\documentclass{article}

\usepackage{arxiv}

\usepackage[utf8]{inputenc} 
\usepackage[T1]{fontenc}    
\usepackage[unicode]{hyperref}
\usepackage{url}            
\usepackage{booktabs}       
\usepackage{amsfonts}       
\usepackage{nicefrac}       
\usepackage{microtype}      
\usepackage{lipsum}		
\usepackage{graphicx}
\usepackage{doi}

\title{Informational Health\\Toward the Reduction of Risks in the Information Space}

\author{ {\hspace{1mm} Fujio Toriumi}\thanks{Use footnote for providing further
		information about author (webpage, alternative
		address)---\emph{not} for acknowledging funding agencies.} \\
	School of Engineering\\
	The University of Tokyo\\
	7-3-1, Hongo, Bunkyo-ku, Tokyo, Japan \\
	\texttt{tori@sys.t.u-tokyo.ac.jp} \\
	\And
	{\hspace{1mm}Tatsuhiko Yamamoto} \\
	Faculty of Law\\
	Keio University\\
	2-15-45, mita, Minato-ku, Tokyo, Japan \\
	\texttt{} \\
}

\renewcommand{\shorttitle}{Informational Health}

\begin{document}
\maketitle

\begin{abstract}
The modern information society, markedly influenced by the advent of the internet and subsequent developments such as WEB 2.0, has seen an explosive increase in information availability, fundamentally altering human interaction with information spaces. This transformation has facilitated not only unprecedented access to information but has also raised significant challenges, particularly highlighted by the spread of ``fake news'' during critical events like the 2016 U.S. presidential election and the COVID-19 pandemic. The latter event underscored the dangers of an ``infodemic,'' where the large amount of information made distinguishing between factual and non-factual content difficult, thereby complicating public health responses and posing risks to democratic processes. In response to these challenges, this paper introduces the concept of ``informational health,'' drawing an analogy between dietary habits and information consumption. It argues that just as balanced diets are crucial for physical health, well-considered nformation behavior is essential for maintaining a healthy information environment. This paper proposes three strategies for fostering informational health: literacy education, visualization of meta-information, and informational health assessments. These strategies aim to empower users and platforms to navigate and enhance the information ecosystem effectively. By focusing on long-term informational well-being, we highlight the necessity of addressing the social risks inherent in the current attention economy, advocating for a paradigm shift towards a more sustainable information consumption model.
\end{abstract}


\section{Introduction}

It has been a long time since modern society came to be recognized as an information society. In particular, the development of the internet has brought about significant changes in the formation of information spaces, unprecedented in human experience. Moreover, the era of WEB 2.0, starting from the mid-2000s, not only enabled people to access information but also accelerated the phenomenon of increasing information quantity through the broadcasting and disseminating of information. Such an increase in information has become a kind of infrastructure supporting our daily lives, from shopping and communicating with friends to reading news articles, dramatically improving convenience.

In this context, the term {\it fake news} became widely known to the general public, triggered by events like the 2016 U.S. presidential election and Brexit. Furthermore, in the context of the COVID-19 pandemic that began in 2020, it became difficult to distinguish between factual and non-factual information due to an overload of information. This state, where it is unclear what to believe or what measures to take, is defined as an ``infodemic." The World Health Organization (WHO) has raised alerts about this, pointing out that the deluge of information has become a problem, akin to the spread of the infectious disease itself\footnote{\url{https://extranet.who.int/kobe_centre/en/tags/infodemic}}. Due to the COVID-19 pandemic, fake news and infodemics not only caused confusion in the information space but also became a serious problem that hindered infectious disease control measures and put people's lives at risk.

With the development of the information society, the amount of information we receive has explosively increased. As a result, while the focus was traditionally on ``how much information we can obtain," now the importance lies in ``how we select and discard information." We use systems like search engines and recommendation systems to assist in the selection of information, but reliance on these systems means entrusting the intake of information to algorithms. The existence of ``filter bubbles," where one encounters only information within the filters set by recommendation systems, has long been pointed out as a cause of the potential loss of information diversity \cite{pariser2011filter}. Similarly, in social media, where one can autonomously select information providers, instead of creating an environment to obtain necessary information, users tend to create a comfortable space with agreeable information. This leads to the formation of ``echo chambers," where one is surrounded by people with similar thoughts and values, resulting in a return of the same opinions and a known loss of information diversity \cite{cinelli2021echo}.

The choice of what information to consume should be left to individual freedom, and the freedom to selectively consume information should also be respected. However, if we are being ``selectively fed" information without knowing its true nature, then we need to secure the right to choose for ourselves. Furthermore, considering how the spread of fake news and infodemics have hindered infection control measures and caused societal chaos, this should be viewed not just as an individual issue but as a problem for society as a whole. The impact of such issues in the information space on democracy itself cannot be ignored.

In response to the various issues inherent in modern information spaces, the authors propose the concept of ``informational health." This paper discusses the risks associated with information spaces and the concept of informational health, as well as what each stakeholder, including users, businesses (platforms, media), and governments, should undertake in response to the new social risks in information spaces, and the technical challenges in realizing this.

\section{Attention Economy and Risks of Information Spaces}
\subsection{Attention Economy}
The attention economy refers to an economic sphere where gaining attention translates into economic incentives and is a fundamental economic principle in modern society \cite{davenport2001attention}. In today's information society, many online platforms operate on an advertising model. Since advertising revenue is gained according to access to content, obtaining as much attention as possible becomes an economic incentive. Furthermore, considering that humans have a limited amount of time available for processing information, a competition arises among services for the limited resource of attention.

Before the internet, media sources such as newspapers, television, and magazines used to provide a bundle of various types of information. They attracted attention by including information likely to draw interest, but they also formatted their offerings in such a way that other pieces of information would catch the eye as well. On the other hand, in the era of internet-based information delivery, content is separated and provided individually. Although many newspapers have their own websites, most users access specific content directly from outside sources.

Furthermore, the current content industry predominantly operates on an advertising-based model. Users can access content for free; however, this is financed through advertising revenues. Users engage with advertisements by viewing them, clicking on them, or making purchases. Through these interactions, the entities managing the content derive financial benefits. This advertising model is not new and has been employed by traditional media such as television and radio. However, in the internet age, the metrics of how often advertisements are displayed, clicked, and lead to purchases are all utilized to generate advertising revenue. Consequently, the number of users viewing the content directly correlates with fluctuations in revenue.

Building on the previously discussed points, the optimal strategy for platforms and media outlets is to attract more attention to individual pieces of content, thereby increasing their viewership. This pursuit of viewership, or high attention, becomes an economic incentive within a market known as the ``attention economy."


It can be argued that the attention economy represents a rational economic principle in contemporary society. In a society where there is a strong perception that ``information is free," content providers find it challenging to generate direct revenue from users. Consequently, they inevitably rely on advertising for income.

On the other hand, it is not a bad thing for users to be offered interesting content for free. Information deemed worthy of attention is made available, and since it is delivered through an advertising model, users do not have to incur any financial costs to access this information. In this sense, the reduction of both search time and financial costs represents a significant advantage for users.

However, platforms overly optimized for the attention economy may prioritize capturing attention and accumulating views at the expense of content quality. In the context of news websites, this might mean that articles are valued not for their societal importance but for their ability to attract views. Additionally, strategies such as using sensational images or headlines to entice clicks — known as ``clickbait" — can be employed to steer users towards articles.

Alternatively, the attention economy might also lead to the loss of the traditional media role of providing a diverse array of information, steering towards a model where only information that aligns with user preferences is offered. This shift could significantly limit the variety of information accessible to the public.

In contemporary society, users also benefit from this situation. For instance, even news articles can be consumed as transient, curiosity-driven content, in which case there is a desire for more engaging content to be provided. Thus, the attention economy brings benefits to users by reducing financial and information gathering costs. However, it also shapes a different information landscape from the traditional one, thereby generating new societal risks.

\subsection{Filter Bubble}

In the attention economy, a primary challenge for platforms is capturing user attention. In this context, recommendation systems are employed. By recommending content that a user is likely to read based on their profile and past browsing history, these systems can increase content access. This not only enhances user engagement for the platform but also benefits users by tailoring and refining the information feed to their specific interests, creating a personalized browsing experience.

However, this is not entirely a positive thing. Algorithms analyze users' preferences and prioritize information based on those preferences, leading to a situation where only information that is deemed interesting to the user is prominently displayed, as if one is trapped inside a bubble of their own making, unable to see anything else but what they want to see. This situation is referred to as a ``filter bubble." Inside this bubble, content and opinions that one likes accumulate, and the exclusion (filtering) of other types makes it difficult to even recognize their existence.

In the attention economy, where users' clicks directly contribute to the platform's revenue, it is more incentivizing for the platform to continue providing information that users are likely to find interesting and click on, rather than delivering a diversity of information. Additionally, as users receive information that interests them, there is no incentive for them to step outside the filter bubble. Therefore, the filter bubble represents a structure that offers significant benefits to platforms, media companies, and users alike.

\subsection{Echo Chamber}
In social networking services (SNS), users tend to form friendships with others who share similar interests and construct their social networks. The desire to communicate with users who share similar values is a natural inclination and can be considered a result of selective exposure. In such environments, one can create an extremely comfortable social space for oneself where shared values are affirmed and one's opinions are more likely to be validated.

However, in such socially constructed spaces, similar to filter bubbles, only information that is of interest or convenient to the individual is visible, while opportunities to encounter a variety of other information are lost. Furthermore, when specific opinions are expressed, an ``echo chamber" is formed where only similar opinions resonate \cite{jamieson2008echo}. This repeated exposure to similar viewpoints can lead individuals to believe that no other opinions exist and that their own are correct without error. In such a state, individuals become vulnerable to misinformation, disinformation, and conspiracy theories. The lack of room for questioning any information can accelerate group polarization. People who are extreme in these tendencies are unable to accept others with differing views and refuse to engage in discussion. This can lead to societal division and poses a threat to democracy.

\subsection{Social Risks Caused by Information Bias}
Filter bubbles and echo chambers introduce biases into the range of information that can be consumed. These biases can pose a risk from an informational perspective in modern society. For example, disinformation, which is deliberately spread to earn advertising revenue by capturing attention, or to tarnish the reputation of celebrities, political groups, or corporations, can be particularly damaging. Even in cases where exposure to diverse information could allow for the discernment of truth from falsehood, biases in information may prevent individuals from questioning the validity of the information presented.

Additionally, the existence of ``social division" based on differences in values has been noted. Social division occurs as a result of exposure only to information aligned with certain values, which can lead to a lack of awareness of other values or an inability to acknowledge them.

In this manner, under the attention economy, it has become difficult to expose oneself in a balanced way to diverse information, resulting in the emergence of numerous societal risks. Nevertheless, within the dominance of the attention economy, the existence of comfortable information spaces holds benefits for both platforms and users, and thus, there is no short-term incentive to depart from them. Users have no incentive to abandon a comfortable information space where interesting information is curated and provided for free. For platforms, there is no benefit in risking a reduction in the number of visits by offering information that users do not prefer. Neither has a need to change their current behavior. In other words, the current information space could be said to be in a Nash equilibrium, from the perspective of game theory. In such a situation, is it possible to change people's information consumption behavior? We must consider how to forego short-term desires and adopt long-term incentives.

\section{Informational Health}
As an example of how we can regulate our short-term desires with long-term incentives, consider our dietary habits. Even in this era of abundance, we can choose to consume a balanced diet for the sake of our physical health, even if it includes ingredients we may not particularly enjoy. Similarly, it is desirable that we can also choose to consume a balanced intake of information, even if it contains information we may not like.

In the era of abundance, our ability to control our diet for long-term incentives is due to a sufficient understanding of the impact of diet on health and the availability of an environment where we can control it ourselves. Similarly, in terms of information, it is necessary to establish an environment where we can control the intake of information that enables us to achieve the ``health" we aim for.

In the information environment, the state in which each individual achieves the ``well-being" they aspire to is called ``informational health."

At this point, let us return to the definition provided by the WHO to consider what constitutes health.

Physical health has been one of the primary concerns for us humans (or any intelligent life form) throughout history. However, ``health" is not solely about the physical aspect. The World Health Organization defines health in the WHO Constitution \cite{world1946health} as follows:

\begin{quote}
HEALTH IS A STATE OF COMPLETE PHYSICAL, MENTAL AND SOCIAL WELL-BEING AND NOT MERELY THE ABSENCE OF DISEASE OR INFIRMITY.
\end{quote}

Namely, the WHO defines health as consisting of three aspects: physical health, mental health, and social health. These aspects of health are interrelated, and the absence of any one of them means that one cannot be considered healthy.

Informational health is deeply related to the three aspects of health. For instance, it is well-remembered that during the COVID-19 pandemic, there was a proliferation of various fake news regarding vaccination. If one cannot consume balanced information, exposure to fake news may impair the ability to make correct decisions about vaccination, potentially harming physical health. Additionally, during the pandemic, there were many reports of deteriorating social relationships due to belief in conspiracy theories, which could negatively affect social and consequently mental health, thereby hindering overall well-being.

To counteract factors that impair our well-being in the information environment, it will be necessary to discuss from the perspective of informational health in the future.

However, the health each aims for is intrinsic, not something externally imposed. Attention must be paid to the fact that the definition of ``being in an informationally healthy state" is not something to be enforced externally.


\section{Technological Development Toward Realization of Informational Health}
\subsection{Dual-Process Theory}
Dual-process theory posits that human thought consists of an implicit and automatic unconscious process (System 1) and an explicit, controlled conscious process (System 2) \cite{wason1974dual}.

We are trained to use the System 2 mode of thought to discern the nutrients necessary for our physical health and to unconsciously avoid dangerous foods through System 1. However, in the recent era of information overload, we are not yet trained to balance the intake of information necessary for informational health. For informational health, it is necessary not only for individuals to change their awareness but also for business efforts and appropriate (yet indirect) support from the government to be provided to create an environment where users can autonomously and proactively select the information they consume. In this paper, we discuss a pathway toward the realization of informational health, taking into account these considerations.

\subsection{Literacy Education to Understand Information Space}
In the modern information space, there are various inhibiting factors that prevent the realization of informational health; hence, the necessity of various forms of environmental development is evident.

Among the most crucial is understanding the nature of the contemporary information space. Just as basic knowledge like ``various nutrients are necessary for health" and ``moderate exercise helps prevent obesity" is effective for physical health, gaining foundational knowledge about the information space is important for achieving individual informational health.

In this context, it is crucial to understand concepts such as the attention economy, filter bubbles, and echo chambers and recognize that we live in such an information space.

The structure of today's digital information space is complex, and even understanding the information environment we are in can be difficult. For example, the information we see is filtered and personalized by recommendation systems—AI, yet less than 20\% of people are aware of this fact \cite{koguchi2023}. This lack of awareness about the information space can lead to susceptibility to conspiracy theories and misinformation, resulting in social division and distrust in media.

Promoting understanding of the digital information space is indispensable, and literacy education plays a key role. While there are concepts like information literacy and media literacy to handle the societal risks posed by information, there is no literacy education specifically for understanding the information space itself. Current literacy education provides guidelines on how to interact with information, such as checking the authors of the information and whether there are alternative opinions, but this places a significant burden on the user. Moreover, without knowledge about the information space itself, even if one acquires literacy skills, putting them into practice is difficult and rarely executed.

Therefore, research and development in literacy education to deepen understanding of the information space itself is necessary. It is important for users to understand ``what situation they are currently in," not just ``what they should do." In Japan, food education has long been practiced, so many Japanese students understand the properties of the foods they consume. For example, from an early age, children learn that vegetables provide vitamins and meat provides protein, and they apply this knowledge to their dietary habits.

Similarly, by acquiring knowledge about information itself, it is expected that transformative changes in information behavior can be encouraged.

\subsection{Visualization of Meta-Information}

\begin{table*}[t]
\centering
\caption{CRAAP Test \cite{blakeslee2004craap}}
\label{tab:CRAAP}
\begin{tabular}{c|l|l} \hline
C&Currency& the timeliness of the information\\ \hline
R&Relevance& the importance of the information for your needs\\ \hline
A&Authority& the source of the information\\ \hline
A&Accuracy& the reliability, truthfulness, and correctness of the content\\ \hline
P&Purpose& the reason the information exists\\ \hline
\end{tabular}
\end{table*}

To achieve informational health, it is important for users to have the opportunity to make proactive and autonomous choices when consuming content and information. However, there are instances where uncritically believing in information selected ``proactively" leads to adherence to conspiracy theories.

We know that maintaining physical health requires keeping a balanced diet on a daily basis. To this end, we use the calorie and nutrient content labels mandated by food labeling laws as one of the criteria for deciding whether to consume a particular food, thereby achieving the intake of health-contributing foods\footnote{Sometimes, physical health may be sacrificed for mental health.}. Similarly, to maintain informational health, the nature and content of information and content must be visualized, and it is considered necessary to use this as a basis for judging what impacts continued consumption of such information may have and what information may be lacking.

Specifically, it would be desirable to make it clear to users what elements and components of content and information are, or how various platforms are displaying and distributing content in balance, so that users can proactively and autonomously decide which content and information to consume for their informational health.

At this time, it is not clear what meta-information affects informational health and in what form. Therefore, it will be necessary to clarify what meta-information should be presented and its impacts in the future.

For example, NewsGuard\footnote{\url{https://www.newsguardtech.com/}} can be considered an example of a meta-information display. It is implemented as a browser plugin, such as in Google Chrome, and provides a service that displays what kind of site is being accessed when browsing news sites, etc. It evaluates the credibility and transparency of each site and can be used as a reference when reading articles.

While NewsGuard evaluates the credibility and transparency of sites, there are many other types of meta-information that could be used to help select content and information. Currently, one of the candidates for meta-information includes criteria considered under media literacy and information literacy.

In information literacy, there is a checklist called the CRAAP test, which highlights five points to be mindful of when acquiring information \cite{blakeslee2004craap}.

By paying attention to these factors, one can achieve high literacy. In terms of the dietary analogy, this corresponds to ``what points to be mindful of to maintain health while eating." This would include knowledge such as balancing the intake of proteins, carbohydrates, and vitamins, avoiding high-calorie items, and researching the origins of foods.

However, while the importance of these checklists can be understood, it is difficult to constantly be aware of all aspects when engaging with information. Constantly worrying about the veracity of information and exploring the intentions behind its dissemination requires a significant burden and cost, and it is certain that the amount of information one can consume would drastically decrease. In other words, it is impossible to constantly consider such meta-information while browsing information content.

Yet, this can be seen as a deficiency in the modern social system. In fact, verifying [the items recommended by these literacies | information through the above-mentioned types of literacy?] on one's own is akin to calculating the vitamin content or calories in a meal or researching its origin by oneself. We decide whether to consume meals based on information provided by producers and sellers. Similarly, for information, it would be desirable not to check these lists ourselves but to have such meta-information, visualized through checklists, provided by media and platform operators. Some of this meta-information could also be automatically presented through technical assistance using artificial intelligence. By visualizing such information, it becomes possible to select the information that should be consumed for informational health.

For instance, if an article has an author's name, one can understand who transmitted it. However, understanding the intentions behind why the author wrote the article would require research on the author. If the author's past articles and positions could be summarized and provided in an easily understandable form, it would facilitate the understanding of the article's content. For example, if it is clear that the person has a deep relationship with a specific company mentioned in the article, one could consider the possibility that the article represents a stakeholder's position while reading. Such assistance can be expected to be realized by using artificial intelligence technology.

\subsection{Informational Health Assessment}
We conduct regular health check-ups to maintain physical health. These allow for the early detection of potential health issues and decision-making regarding necessary treatments.

Similarly, it would be beneficial to provide users with regular opportunities for an informational health assessment to check their own informational health. By voluntarily participating in such information exams, individuals can visualize what kind of information they have been exposed to, which can motivate changes in their information consumption behaviors.

In an informational health assessment, it could be possible to analyze the diversity of the information sources being consumed (to see if there is any bias) and the credibility of those sources and to present these results as objective data. Additionally, the extent to which one is ``afflicted" by filter bubbles or echo chambers could be made visible.

Furthermore, it would be desirable to develop and provide services that visualize statistical information about the information one has previously encountered and how information is presented to other users, thus visualizing one's position within the information space. The recent advancements in natural language and image processing technologies are likely to significantly contribute to the development of these societal systems.

\subsection{Incentive for Platforms}

\begin{table}[bt]
\centering
\caption{Changes in User Retention: Diverse Article Viewing (Intervention Group) vs. Homogeneous Article Viewing (Control Group)} \label{tab:ATT micro diversity retention}
\begin{tabular}{|c|c|c|}
\hline
\begin{tabular}[c]{@{}c@{}}\end{tabular} & 2019 & 2021 \\ \hline
Effect of Intervention & \textbf{+0.28***} & \textbf{+0.17***} \\ \hline
Standard Error & 0.0040 & 0.0071 \\ \hline
\end{tabular}
\end{table}

When users desire exposure to diverse information, platforms can implement algorithms that provide a variety of content and help avoid filter bubbles. However, if such an algorithm leads to a decrease in viewership, considering the principles of the attention economy, it may result in a loss of economic incentives. Therefore, under the current system of the attention economy, there is no short-term incentive for platform operators to change their practices.

On the other hand, several studies suggest that exposure to diverse information can encourage users to continue using the same platform for a longer period. For example, analyses conducted on the music streaming service Spotify \cite{anderson2020algorithmic} and on a recommendation platform with billions of users \cite{google2022} have shown that users exposed to diverse information are more likely to continue using the platform.

Table \ref{tab:ATT micro diversity retention} presents the results of our analysis on the diversity of articles viewed and the continuation rates of users for the years 2019 and 2021, conducted with the cooperation of Gunosy Inc. The comparison of continuation rates between users who viewed diverse articles and those who viewed homogeneous articles was performed using propensity score matching \cite{rosenbaum1983central}. These results clearly demonstrate that users who viewed a variety of articles tended to continue using the service. Thus, from the perspective of continuity, it is evident that the presence of users who view diverse articles can provide a long-term incentive for sustained use for platform operators.

These examples suggest that the common belief that providing diverse articles leads to a lack of economic incentives is not always correct according to the data. Similarly, by understanding the information space, it may be possible to implement measures that offer incentives for platform operators, media, users, and society to achieve informational health.

Additionally, there are recommendation algorithms that increase diversity without reducing click rates, indicating that increasing diversity does not necessarily lead to a decrease in platform incentives \cite{sonoda2024}.

\section{Conclusion}
The attention economy, which dominates the modern information space, has the potential to introduce numerous social risks. On the other hand, both users and platforms are currently stuck in a local minimum, lacking incentives for improvement, making resolution difficult. In this context, this study focuses on the relationship between diet and health, noting that incentives for long-term health can sometimes outweigh short-term desires. It proposes the concept of informational health by using the relationship between diet and physical health as an analogy.

Informational health does not contradict the health definition provided by the WHO but can play a role in it. This paper demonstrates the potential of informational health as a new form of health. To realize informational health, we proposed three strategies: literacy education to understand the information space, visualization of meta-information, and informational health assessment. We also show that users who browse a variety of information have higher retention. This means that providing information with consideration for informational health has certain value to platforms as well.

Our own well-being is ours alone and should not be compromised by anyone. In this sense, the current information space is in a state where informational health is being compromised by various forms of information hacking. The goal of reclaiming our informational health is expected to greatly benefit from artificial intelligence technologies such as natural language processing and behavioral analysis.

On the other hand, achieving a healthy discourse space, which is the goal of informational health, would require changes not only from platforms, the media, telecommunications carrier, and advertisers but also from users themselves. Furthermore, various supportive measures from the government will likely be necessary. The concept of informational health has been discussed in the Ministry of Internal Affairs and Communications' research group on ``Platform Services\footnote{\url{https://www.soumu.go.jp/main_sosiki/kenkyu/platform_service/index.html}}" and ``Ensuring the Integrity of Information Circulation in Digital Spaces\footnote{\url{https://www.soumu.go.jp/main_sosiki/kenkyu/digital_space/index.html}}." It is also highlighted in the Ministry of Internal Affairs and Communications' 2023 White Paper on Information and Communications in Japan \cite{whitepaper2023}. It has already been covered by many media outlets and has gained a certain level of recognition in Japan.

To achieve a healthy discourse space, it will be necessary to involve even more stakeholders and discuss what the information environment should be like. In this context, the concept of informational health is considered to be useful.



\bibliographystyle{unsrt}


\begin{thebibliography}{10}

\bibitem{pariser2011filter}
Eli Pariser.
\newblock {\em The filter bubble: What the Internet is hiding from you}.
\newblock penguin UK, 2011.

\bibitem{cinelli2021echo}
Matteo Cinelli, Gianmarco De~Francisci~Morales, Alessandro Galeazzi, Walter Quattrociocchi, and Michele Starnini.
\newblock The echo chamber effect on social media.
\newblock {\em Proceedings of the National Academy of Sciences}, 118(9):e2023301118, 2021.

\bibitem{davenport2001attention}
Thomas~H Davenport and John~C Beck.
\newblock The attention economy.
\newblock {\em Ubiquity}, 2001(May):1--es, 2001.

\bibitem{jamieson2008echo}
Kathleen~Hall Jamieson and Joseph~N Cappella.
\newblock {\em Echo chamber: Rush Limbaugh and the conservative media establishment}.
\newblock Oxford University Press, 2008.

\bibitem{world1946health}
WHO.
\newblock Health is a state of complete physical, mental and social well-being and not merely the absence of disease or infirmity.
\newblock In {\em International Health Conference, New York}, pages 19--22, 1946.

\bibitem{wason1974dual}
Peter~C Wason and J~St~BT Evans.
\newblock Dual processes in reasoning?
\newblock {\em Cognition}, 3(2):141--154, 1974.

\bibitem{koguchi2023}
Teppei Koguchi and Toshiya Jitsuzumi.
\newblock Analysis of the acceptability of the" information health" concept: Consideration of the possibility of beneficiary payment.
\newblock International Telecommunications Society (ITS), 2023.

\bibitem{blakeslee2004craap}
Sarah Blakeslee.
\newblock The craap test.
\newblock {\em Loex Quarterly}, 31(3):4, 2004.

\bibitem{anderson2020algorithmic}
Ashton Anderson, Lucas Maystre, Ian Anderson, Rishabh Mehrotra, and Mounia Lalmas.
\newblock Algorithmic effects on the diversity of consumption on spotify.
\newblock In {\em Proceedings of The Web Conference 2020}, pages 2155--2165, 2020.

\bibitem{google2022}
Yuyan Wang, Mohit Sharma, Can Xu, Sriraj Badam, Qian Sun, Lee Richardson, Lisa Chung, Ed~H. Chi, and Minmin Chen.
\newblock Surrogate for long-term user experience in recommender systems.
\newblock In {\em Proceedings of the 28th ACM SIGKDD Conference on Knowledge Discovery and Data Mining}, KDD '22, page 4100--4109, New York, NY, USA, 2022.

\bibitem{rosenbaum1983central}
Paul~R Rosenbaum and Donald~B Rubin.
\newblock The central role of the propensity score in observational studies for causal effects.
\newblock {\em Biometrika}, 70(1):41--55, 1983.

\bibitem{sonoda2024}
Atom Sonoda, Fujio Toriumi, and Hiroto Nakajima.
\newblock User experiments on the effect of the diversity of consumption on news services.
\newblock {\em IEEE Access}, 2024.

\bibitem{whitepaper2023}
{Ministry~of~Internal~Affairs~and~Communications}.
\newblock 2023 white paper information and communications in japan.
\newblock 2023.

\end{thebibliography}





\end{document}